# Nanoparticle Radiosensitization: from extended local effect modeling to a survival modification framework of compound Poisson additive killing and its carbon dots validation

Hailun Pan [1,2], Xiaowa Wang [1,2,3], Aihui Feng [1,2,4], Qinqin Cheng [1,2], Xue Chen [5], Xiaodong He [6,10], Xinglan Qin [1,2], Xiaolong Sha [1,2], Shen Fu [2,7,8], Cuiping Chi [9] and Xufei Wang [1,2,10]

1. Institute of Modern Physics, Fudan University, Shanghai 200433, China
2. Key Laboratory of Nuclear Physics and Ion-beam Application (MOE), Fudan University, Shanghai 200433, China
3. Shanghai Proton and Heavy Ion Center, Shanghai 201321, China
4. Radiotherapy Department, Shanghai Chest Hospital, Shanghai Jiao Tong University, Shanghai 200025, China
5. Department of Radiation Oncology, Fudan University Shanghai Cancer Center, Shanghai, 200032, China
6. Radiotherapy Department, Ruijin Hospital, Shanghai Jiao Tong University, Shanghai 200025, China
7. Proton & Heavy Ion Medical Center, State Key Laboratory of Radiation Medicine and Protection, School of Radiation Medicine and Protection, Soochow University, Suzhou, 215006, China.
8. Department of Radiation Oncology, Shanghai Concord Cancer Hospital, Shanghai 200020, China.
9. Department of Radiation and Environmental Medicine, China Institute for Radiation Protection, Taiyuan, 030006, China
10. Author to whom correspondence should be addressed.

E-mail: wangxufei@fudan.edu.cn; xiaodonghe888@sina.cn

## Abstract

*Objective*: To construct an analytical model instead of local effect modeling for the prediction of the biological effectiveness of nanoparticle radiosensitization. *Approach*: An extended local effects model is first proposed with a more comprehensive description of the nanoparticles mediated local killing enhancements, but meanwhile puts forward challenging issues that remain difficult and need to be further studied. As a novel method instead of local effect modeling, a survival modification framework of compound Poisson additive killing is proposed, as the consequence of an independent additive killing by the assumed equivalent uniform doses of individual nanoparticles per cell under the LQ model. A compound Poisson killing (CPK) model based on the framework is thus derived, giving a general expression of nanoparticle mediated LQ parameter modification. For practical use, a simplified form of the model is also derived, as a concentration dependent correction only to the α parameter, with the relative correction ($\alpha''/\alpha$) dominated by the mean number, and affected by the agglomeration of nanoparticles per cell. For different agglomeration state, a monodispersion model of the dispersity factor $\eta=1$, and an agglomeration model of $2/3<\eta<1$, are provided for practical prediction of ($\alpha''/\alpha$) value respectively. *Main results*: Initial validation by the radiosensitization of HepG2 cells by carbon dots showed a high accuracy of the CPK model. In a safe range of concentration (0.003-0.03 μg/μL) of the carbon dots, the prediction errors of the monodispersion and agglomeration models were both within 2%, relative to the clonogenic survival data of the sensitized HepG2 cells. *Significance*: The compound Poisson killing model provides a novel approach for analytical prediction of the biological effectiveness of nanoparticle radiosensitization, instead of local effect modeling.
.

# Contents

# 1. Introduction

Nanoparticle radiosensitization has been intensively studied in last decade as an effective method for radiotherapy enhancement, with great progresses achieved on both the radiosensitizer materials from high-Z metallic nanoparticles (Schuemann et al 2020, Penninckx et al 2020, Sajo and Zygmanski 2020, Kempson 2021) to metallic/nonmetallic "sensitization plus" nanoagents for a multifunctional theranostics (Bilynsky et al 2021, Yi et al 2021, Denkova et al 2020, Thorat and Bauer 2019, Yan et al 2021), and a deeper understanding of the radiosensitization mechanism from the subcellular local dose enhancements that were widely investigated by *in silico* macro- to micro-dosimetry simulations (Moradi et al 2021, Vlastou et al 2020, Peukert et al 2020a) to the local-to-global physicochemical (Verkhovtsev et al 2015, Haume 2018, Rudek et al 2019, Hespeels et al 2019, Peukert et al 2019, 2020b), biochemical and biological enhancements (Chen et al 2020, Liu et al 2020, Sun et al 2020), including enhanced reactive oxygen species and oxidative stress (Butterworth et al 2012, 2013, Klein et al 2014, 2018, Howard et al 2020, Clement et al 2020, Jia et al 2021), cell cycle arrest (Liu et al 2015, Rieck 2019, Bromma et al 2019, Abbasian et al 2019), endoplasmic reticulum stress (Yasui et al 2014, Zhu et al 2018), DNA repair inhibition (Turnbull et al 2019), etc. that were revealed in a lot of *in silico, in vitro* and *in vivo* studies. Based on the material development, dosimetric simulation and biological understandings, preclinical model of the biological effectiveness (McMahon et al 2011, 2016a, Lechtman et al 2013, Zygmanski et al 2013, Lin et al 2014, 2015, Lechtman and Pignol 2017, Ferrero et al 2017, Sung et al 2017, 2018, Sung and Schuemann 2018, Brown and Currell 2017, Villagomez-Bernabe and Currell 2019, Villagomez-Bernabe et al 2021, Kim et al 2021, Engels et al 2020, Batooei et al 2021, Hahn and Zutta Villate 2021, Melo-Bernal et al 2018,2021), for the planning of nanoparticles enhanced radiotherapy, also made essential progresses. Clinical trials also reported promising (Bonvalot et al 2019a, Verry et al 2019, 2021, Hoffmann et al 2021, Bort et al 2020) and somewhat controversial (Vilotte et al 2019, Bonvalot et al 2019b)) initial results.

However, for a clinical translation of nanoparticles enhanced radiotherapy, the *in vitro*, *in vivo* and *in silico* knowledges to date are still insufficient for constructing a comprehensive dosimetric model for the nanoparticles mediated biological effectiveness, due to the modeling difficulties for the highly inhomogeneous subcellular enhancement effects. In the reported modeling studies (as mentioned above), at least three issues related to the local effects need to be further addressed but still challenging: (1) Most studies performed Monte Carlo simulations of the local dose of secondary electrons, from high-Z metallic nanoparticles (MNPs) artificially located in a cell model, with the nucleus seen as the only sensitive volume for radiation killing. Clearly, the biological effectiveness predicted in such models is highly dependent on the specific subcellular "source-target" geometry, and incomplete by ignoring both the local dose enhanced killing via extranulcear organelles (e.g. mitochondria, endoplasmic reticulum, plasma membrane) and the killing enhancement due to the local-to-global biochemical effects (Chen et al 2020, Liu et al 2020, Howard et al 2020). In fact, the often visibly higher experimental sensitizations (Butterworth et al 2012, Ruan et al 2018, Delorme et al 2017, Sotiropoulos et al 2017) in comparison to the simulated nucleus dose enhancements, and the directly observed biological consequences (Forkink et al 2010, Taggart et al 2014, Štefančíková et al 2016, McQuaid et al 2016, Ghita et al 2017, Pagáčová et al 2019), all strongly indicated the physical and/or bio-chemical killing contributions via extranuclear organelles. (2) Though there have been some simulation studies reporting the MNPs mediated local dose to the extranuclear targets like mitochondria (Kirkby and Ghasroddashti 2015, McNamara et al 2016, McMahon et al 2016b, Zein 2017, Francis et al 2019, Hahn and Zutta Villate 2021), highlighting the killing contribution by the local dose deposited in cytoplasm, a prediction of the non-DNA target killing still failed, due to the lack of knowledge on the dose killing relationship through the cytoplasm targets; (3) Though many types of (high-Z/low-Z metallic/nonmetallic) nanoparticles have demonstrated a series of biochemical mechanisms that lead to enhanced radiosensitivity, to the authors' knowledge, there is so far no modeling methods for the enhanced killing by the local-to-global biochemical effects. In summary, these issues in the modeling of biological effectiveness make the treatment planning for nanoparticle enhanced radiotherapy still a challenging task.

To advance the clinical practice of nanoparticle radiosensitization, in this work, an extended local effects model is first proposed (section 2.2), containing the local dose enhanced killing in both nucleus and cytoplasm, and local-to-global biochemical effects. However, quantification of the nanoparticle enhanced killing is still challenging, due to the fundamental issues in the subcellular local effect modeling. To avoid the issues, a survival modification framework of compound Poisson additive killing, instead of local effect modeling, is constructed for an analytical description of nanoparticles mediated radiosensitization (section 2.3-2.4), and validated in vitro by the clonogenic dose survival of HepG2 cells sensitized via a common type of carbon dots (section 3).

## 2. Theoretical framework

### 2.1 General survival model of intrinsic radiation killing

According to the target theory in radiobiology, in an arbitrary single cell exposed to a homogenous radiation dose $D$, the number of lethal damages (killing events) $l$ follows a binomial distribution, and approaches a Poisson distribution

$$p(l|D) = \frac{\lambda(D)^l}{l!} e^{-\lambda(D)} \tag{1}$$

when the incident photon or particle number of dose $D$ approaches infinity. $\lambda(D)$ is dose-dependent mean number of intracellular radiation killing events. Therefore the dose-survival probability is the probability that a single cell underwent zero killing events after an exposure to radiation dose $D$

$$S(D) = p(l=0|D) = \frac{\lambda(D)^0}{0!} e^{-\lambda(D)} = e^{-\lambda(D)} \tag{2}$$

It is known that for most mammalian cells exposed to a homogenous radiation dose, the mean number of Poisson killing events per cell can be described by the Linear-Quadratic (LQ) relationship

$$\lambda(D) = -\ln S(D) = \alpha D + \beta D^2 \tag{3}$$

that reflecting the intrinsic dose killing contributed from the entire cell. For cytoplasm targets like mitochondria, a much higher radioresistance leads to a far lower density of extranuclear killing events under the same dose, thus in most radiobiology models, nucleus is seen as the unique organelle for radiation killing, through the lethal DNA damage via single and double string breaks.

In a more general form of the intrinsic dose survival, assuming that the initial damages occurring in nucleus and cytoplasm are induced separately (ignoring the possible downstream synergies), according to Raikov's theorem, the two parts of killing events respectively follow the Poisson distributions

$$p_\text{n}(l|D) = \frac{\lambda_\text{n}(D)^l}{l!} e^{-\lambda_\text{n}(D)}, \quad p_\text{cy}(l|D) = \frac{\lambda_\text{cy}(D)^l}{l!} e^{-\lambda_\text{cy}(D)} ; \quad \lambda_\text{n}(D) + \lambda_\text{cy}(D) = \lambda(D) \tag{4}$$

where $\lambda_n(D)$, $\lambda_{cy}(D)$ are the mean numbers of killing events in nucleus and cytoplasm, respectively. Due to the heterogeneity of radiosensitivity in subcellular domains, the dose-dependent killing events in a whole cell is dominated by intranuclear killing, thus equation (4) can be rewritten as

$$\begin{cases} p_n(l|D) \approx p(l|D) = \dfrac{(\alpha D + \beta D^2)^l}{l!} e^{-(\alpha D + \beta D^2)}, & \lambda_n(D) \approx \lambda(D) = \alpha D + \beta D^2 \\ p_{cy}(l|D) = \dfrac{\nu(D)^l}{l!} e^{-\nu(D)}, & \lambda_{cy}(D) = \nu(D) \ll \alpha D + \beta D^2 \end{cases} \quad (5)$$

and the dose-survival probability of a single cell, as given in equations (2) and (3), can be expressed in a more generalized form

$$S(D) = p(l=0|D) = p_n(l=0|D) \times p_{cy}(l=0|D) = e^{-(\alpha D + \beta D^2)} \times e^{-\nu(D)} \approx e^{-(\alpha D + \beta D^2)} \quad (6)$$

## 2.2 Nanoparticles mediated survival modification: local effect modeling and fundamental issues

### 2.2.1 Enhanced killing leads to survival modification

What are we talking about when we talk about nanoparticles radiosensitization? From a physicist's viewpoint, in most scenarios especially in vitro cases, such a phenomenon can be described as a nano-exposure mediated radiobiological consequence, where randomly and agglomeratively distributed nanoparticles in cytoplasm (and sometimes pericellular) lead to a clonogenic survival modification (reduction) of the irradiated cells, through the increased radiation killing events by the nanoparticle mediated local dose enhancement and/or local-to-global biochemical effects (including enhanced radiolysis yield, e.g. reactive oxygen species). Therefore, to make a prediction of the cell survival modification, a quantitative model of increased killing events per cell needs to be established, based on the essential features of the subcellular interactions mediated by the intra/peri cellular nanoparticles.

### 2.2.2 An extended local effects model of enhanced killing

To establish a quantitative model of the increased killing events per cell, a more comprehensive description of the nanoparticles mediated interactions is needed. By incorporating the local dose effects in cytoplasm and the local-to-global biochemical effects across a whole cell, an triple local effects model (TLEM), is proposed as a conceptual whole-cell effect model for nanoparticle radiosensitization, with the sub-models called TLEM-I, II, III, for the intranuclear local dose effect, the intracellular local dose effects and the whole-cell local dose plus local-to-global biochemical effects, respectively, as illustrated in figure 1.

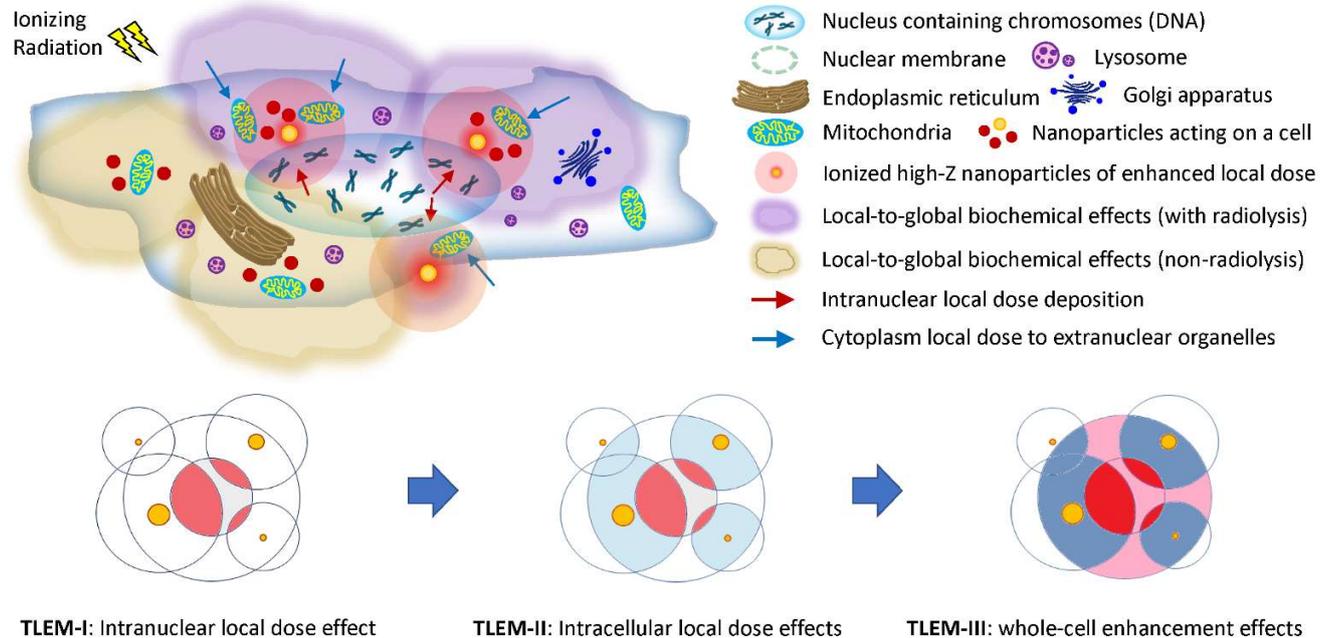

**Figure 1**. Conceptual illustration of the extended local effects model TLEM for nanoparticle mediated radiosensitization. At least three types of nanoparticle-mediated interactions are exerted to a target cell (upper part): intranuclear local dose enhancement, intra-cytoplasm local dose enhancement and local-to-global biochemical enhancements. The TLEM is proposed with the sub-models called TLEM-I, II, III (lower part), for the intranuclear local dose effect, the intracellular local dose effects, and the whole-cell local dose plus biochemical effects, respectively.

According to the concept of TLEM, nanoexposure leads to uptake and (in most cases) cytoplasm accumulation of nanoparticles, and in cell irradiation, the subcellular interactions mediated by the cytoplasm-located (and unwashed pericellular) nanoparticles are divided as three parts (figure 1): the local dose enhancements in nucleus and cytoplasm respectively, and the local-to-global biochemical effects. The local dose is contributed by secondary electrons deposited outside the volume of high-Z nanoparticles, while the biochemical effects originate from nanoparticle surface, but may lead to global consequence throughout a whole cell. Noted that the TLEM does not mean that the triple effects must coexist. Generally most kinds of nanoparticles can induce some different levels of intracellular biochemical effects (e.g. oxidative stress, cycle arrest or surface modification mediated reactions) that lead to an enhanced radiosensitivity, while high-Z nanoparticles will introduce an additional, or even a dominant local dose enhanced killing, in both nucleus and cytoplasm.

It can be seen that nanoparticles radiosensitization is strongly characterized by its multiple inhomogeneities in subcellular scale: (1) the inhomogeneity of radiosensitivity in subcellular domains, (2) the inhomogeneity of distribution of nanoparticles in the subdomains, and (3) the inhomogeneity of a single nanoparticle's local action on the surroundings. These triple inhomogeneities, together lead to inhomogeneous distributions of the triple local enhancements in different subdomains of a single cell. In TLEM, two subdomains, nucleus and cytoplasm, are defined in a single cell model with distinctly different radiosensitivity for the local dose effects. The distribution of nanoparticles in the subdomains is defined as a random agglomerative distribution in cytoplasm, plus some amount of pericellular particles depending on the post-nanoexposure treatments (washing or not), but none in nucleus. A single nanoparticle's local enhancement, as a function of local action on the surroundings, contains not only the high gradient dose of secondary electrons emitting from the nanoparticle surface, but also the biochemical interactions exerted on surrounding medium or organelle targets through contact reactions or surface catalysis.

Clearly, TLEM gives a better description on the core features of nanoparticles mediated subcellular "source-target" interactions. With the multiple inhomogeneities defined in TLEM, in principle, the increased number of killing events per cell that leads to the survival modification can be quantified, using a spatial integral of the killing density throughout the volume of a whole cell. However, the TLEM killing quantification still needs more specific features and parameters to be defined, as discussed in 2.2.3.

### 2.2.3 Survival modification under the extended model – challenges in quantification

To make a quantitative modeling of the increased killing events per cell, the TLEM needs more explicit features and parameters: (1) carefully defined geometries of the nucleus and cytoplasm, (2) the distinct radiosensitivities of the nucleus and the cytoplasm, (3) a model of single nanoparticle's local action and its killing densities in nucleus and cytoplasm, respectively, and (4) a spatial distribution model of the nanoparticles in cytoplasm (plus pericellular).

However, further definition of the features and parameters puts forward fundamental issues for the TLEM killing quantification: (i) What is the radiosensitivity of a cytoplasm? To predict the local dose enhanced killing in cytoplasm, the intrinsic dose killing density in cytoplasm needs to be determined, but to our knowledge very limited data is available for the dose-killing relationship via the subdomain of cytoplasm as a whole, though subcellular microbeams have been widely used in basic radiobiology studies (Ghita et al 2018). It should be noted that, though the weight of cytoplasm in radiation cell killing is far less than that of nucleus, in the radiosensitization by high-Z nanoparticles, the enhanced local dose deposited around the nanoparticles may be very high (Rabus et al 2019), leading to nonignorable, even severe lethal damage to closely contacted subsensitive organelles in cytoplasm (e.g. mitochondria or endoplasmic reticulum). These enhanced killing events may lead to a modification of the cell dose survival, even a nonlinear deviation (Rashid et al 2018) from the intrinsic LQ behavior dominated by DNA damages. In brief, the intrinsic radiosensitivity of cytoplasm is essential for the TLEM modeling but still unknown and difficult to measure by existing methods. (ii) How to quantify the local-to-global biochemical enhancements ? Different from the local dose enhanced killing that can be modeled by Monte Carlo method, the biochemical enhancement, with both the local damage like mitochondrial oxidative stress and the global consequence like cell cycle arrest, is still difficult to model for the lack of methods. (iii) Moreover, the geometries of subcellular domains, the spatial distribution and agglomeration effect of the nanoparticles, all need finer modeling to improve the prediction accuracy, but are still challenging for the local effects that highly dependent on specific "source-target" geometry, due to the multiple subcellular inhomogeneities.

As seen from the above discussion, to make a reliable prediction of the survival modification in nanoparticles radiosensitization, the core task is to build a quantification model for the probability distribution of the increased number of killing events per cell. The triple local effects model proposes an extended quantification of whole cell killing events from both nucleus and cytoplasm, containing both local dose and biochemical killing effects, thus to make a complete modeling of survival modification. However, to make a practical quantification by the TLEM, there are still many fundamental issues to be solved. To avoid these difficulties, in the following part, an alternative statistical model is derived, for an analytical quantification of nanoparticle enhanced killing.

## 2.3 A survival modification framework of compound Poisson additive killing

### 2.3.1 Basic framework

If we assume an independent occurrence of intrinsic radiation killing and nanoparticle mediated exogenous killing enhancement, similar to equations (4)-(6), a general formula of nanoparticles mediated dose survival modification can be written as

$$S(D,C) = P(l=0|D) \times P(L_Z=0|D,C) = e^{-(\alpha D + \beta D^2)} e^{-\nu(D)} \gamma(D,C) \approx e^{-(\alpha D + \beta D^2)} \gamma(D,C) \quad (7)$$

$P(l=0|D)$ is the intrinsic probability of zero killing (survival) in a single cell under the radiation dose $D$, $L_Z$ is the killing events added by nanoparticles, $P(L_Z=0|D,C)$ is the probability of zero killing by the nanoparticles under the dose $D$ and concentration $C$, thus $S(D,C)$ is the dose survival probability of a single cell modified by the nanoparticles at the concentration $C$. (Note that $C$ refers to the applied nanoparticle solution concentration for cell treating, not the intracellular value of absorbed nanoparticles). The intrinsic zero killing (survival) probability $P(l=0|D)$ follows the LQ model, as seen in equation (7), while the probability of zero killing mediated by the nanoparticles is denoted as a modification function $\gamma(D,C)$ acting on the intrinsic dose survival

$$P(L_Z=0|D,C) = \gamma(D,C) \quad (8)$$

To model the probability distribution of the increased killing events per cell, as a homogenization approximation regardless of the inhomogeneities of the subcellular nanoparticle enhancements, two basic quantities can be assumed to be Poisson variables: (1) The number of nanoparticles per cell $Z$, i.e. the number of (absorbed and pericellular) nanoparticles acting on each cell, that following a Poisson distribution $Z \sim \text{Pois}(\bar{\lambda})$; (2) The number of killing events per nanoparticle, $L_1$, i.e. the killing events number increased by a single nanoparticle in a cell, which following a Poisson distribution $L_1 \sim \text{Pois}(\mu_1)$. For specific types of cells and nanoparticles, the mean number of nanoparticles per cell $\bar{\lambda}$ should have a dependency on the applied nanoparticle concentration $\bar{\lambda}=\bar{\lambda}(C)$, while the mean number of increased killing events per nanoparticle should have a binary dependency of $\mu_1=\mu_1(D,C)$, on both the applied nanoparticle concentration $C$ and the radiation dose $D$. The two Poisson random variables are expressed as

$$\begin{cases} Z \sim \text{Pois}(\bar{\lambda}) \to P(Z=z) = \frac{\bar{\lambda}^z}{z!} e^{-\bar{\lambda}}, & E(Z) = V(Z) = \bar{\lambda} = \bar{\lambda}(C) \\ L_1 \sim \text{Pois}(\mu_1) \to P(L_1=l) = \frac{\mu_1^l}{l!} e^{-\mu_1}, & E(L_1) = V(L_1) = \mu_1 = \mu_1(D,C) \end{cases} \quad (9)$$

Based on the two Poisson distributions and a further assumption of independent occurrence of the killing events per nanoparticle, the total killing events by all the nanoparticles per cell can be described as an additive killing of a compound Poisson distribution

$$\begin{aligned} P(L_Z=l|D,C) &= P\left(\sum_{i=1}^{Z} L_{1i} = l\right) = \sum_{n=0}^{\infty} P\left(\sum_{i=1}^{n} L_{1i} = l\right) P(Z=n) \\ &= \sum_{n=0}^{\infty} \left[\left(\frac{\bar{\lambda}^n}{n!} e^{-\bar{\lambda}}\right) P\left(\sum_{i=1}^{n} L_{1i} = l\right)\right] = \sum_{n=0}^{\infty} \left[\frac{(n\mu_1)^l}{l!} e^{-n\mu_1}\right] \left(\frac{\bar{\lambda}^n}{n!} e^{-\bar{\lambda}}\right) \end{aligned}, \quad \begin{pmatrix} E(L_Z) = \bar{\lambda}\mu_1 \\ V(L_Z) = \bar{\lambda}\mu_1(1+\mu_1) \\ \bar{\lambda}=\bar{\lambda}(C), \ \mu_1=\mu_1(D,C) \end{pmatrix} \quad (10)$$

Thus the probability of zero killing added by all the nanoparticles per cell (i.e. the survival modification function) is

$$P(L_Z=0|D,C) = \gamma(D,C) = \sum_{n=0}^{\infty} e^{-n\mu_1(D,C)} \left(\frac{\bar{\lambda}(C)^n}{n!} e^{-\bar{\lambda}(C)}\right) \quad (11)$$

Substitute equation (11) into (7), the cell dose survival under the compound Poisson additive killing is modified as

$$S(D,C) = P(l=0|D) \times P(L_Z=0|D,C) = e^{-(\alpha D + \beta D^2)} \sum_{n=0}^{\infty} e^{-n\mu_1(D,C)} \left(\frac{\bar{\lambda}(C)^n}{n!} e^{-\bar{\lambda}(C)}\right) \quad (12)$$

It can be seen that under the statistical framework of compound Poisson additive killing, the nanoparticles mediated cell survival modification is dependent on both the applied concentration $C$ and the radiation dose $D$.

### 2.3.2 $\bar{\lambda}(C)$: using the mean number of nanoparticles per cell for a Poisson approximation

As a first-order approximation, the number of nanoparticles per cell can be simplified from the Poisson number $Z \sim \text{Pois}(\bar{\lambda}(C))$ to a deterministic quantity of the mean nanoparticle number per cell $Z=\bar{\lambda}(C)$, while the killing events number per nanoparticle remains the Poisson variable $L_1 \sim \text{Pois}(\mu_1(D,C))$. Such a simplification degrades the compound Poisson killing in equation (10) to a Poisson distribution by a deterministic number of nanoparticles, on the basis of the additive property of Poisson distribution

$$P(L_Z = l | D, C) = \frac{\left(\bar{\lambda}(C)\mu_1(D,C)\right)^l}{l!} e^{-\bar{\lambda}(C)\mu_1(D,C)} \tag{13}$$

Correspondingly, the zero killing probability (i.e. the survival modification function) is simplified from equation (11) to

$$P(L_Z = 0 | D, C) = \gamma(D, C) = e^{-\bar{\lambda}(C)\mu_1(D,C)} \tag{14}$$

thus the cell dose survival under the approximated Poisson killing is simplified from equation (12) to

$$S(D, C) = e^{-(\alpha D + \beta D^2)} e^{-\bar{\lambda}(C)\mu_1(D,C)} \tag{15}$$

### 2.3.3 $\bar{d}_1(D, C)$: define an equivalent dose per nanoparticle for killing events quantification

To have an operable quantification model for the nanoparticles mediated killing events under the compound Poisson framework, the Poisson mean of the number of killing events per nanoparticle, $\mu_1(D,C)$, requires an explicit expression. Unlike the TLEM quantification that needs to track down and sum over the local killing events in the whole volume of a cell, an equivalent global quantity, called a single nanoparticle's equivalent dose $\bar{d}_1$, is introduced to quantify the physical and biochemical killing events by a single nanoparticle to a whole cell. For this, $\bar{d}_1$ is defined as a hypothetical uniform global dose in a whole cell to account for the assumed Poisson killing events $L_1 \sim \text{Pois}(\mu_1(D,C))$, therefore under the LQ model of radiation killing, the Poisson mean of the killing events per nanoparticle $\mu_1$ can be expressed as a LQ function of the equivalent dose $\bar{d}_1$

$$\mu_1(D, C) = \alpha \bar{d}_1(D, C) + \beta \bar{d}_1^{\,2}(D, C) \tag{16}$$

The introduction of single nanoparticle' equivalent dose has three assumptions: (1) The stochastic biochemical effect of a single nanoparticle is equated to a Poisson killing by a deterministic radiation dose following the LQ model, (2) The subcellular local physical and biochemical effects of a single nanoparticle, are equated to a global effect of a uniform dose over a whole cell, (3) Corresponding to the independent killing by different nanoparticles assumed for equation (10), the collective effect of multiple nanoparticles in cell killing, is equated to an additive killing by multiple equivalent doses without the square cross terms under the LQ model. That means the increased killing events per cell by the subcellular physical and/or biochemical effects of multiple nanoparticles, is equated to the global effect of additive radiation killing by multiple independent exposures to the homogeneous equivalent dose $\bar{d}_1$. With the expression through $\bar{d}_1$ (equation (16)), the nanoparticles enhanced killing per cell can be quantified under the LQ model. Therefore, the following work turns to finding the function relationship of $\bar{d}_1(D, C)$, as discussed in 2.3.4.

### 2.3.4 $\bar{\lambda}(C)$ and $\bar{d}_1(D, \bar{\lambda}(C))$: construction of two nonlinear dependencies

So far the Poisson mean of the number of nanoparticles per cell $\bar{\lambda}(C)$ and the single nanoparticle' equivalent dose $\bar{d}_1(D, C)$ are both still unknown functions. For $\bar{\lambda}(C)$, there is a lot of data (Chithrani et al 2006, Kalambur et al 2007, Tahara et al 2009, Jain et al 2011, Lunov et al 2011, Jiang et al 2013) demonstrating a saturation behavior of cellular uptake of nanoparticles with the increase of treating concentration, therefore, a negative exponential function can be used to describe the nonlinear dependency

$$\bar{\lambda}(C) \approx A_1 \left(1 - e^{-k_1 C}\right) \tag{17}$$

where $A_1$ represents the mean saturation number of nanoparticles per cell, $k$ is the rate constant with the dimension of reciprocal concentration. The $A'$, $k$ parameter values rely on specific cells, nanoparticles and co-incubation conditions. From equation (17) the dependency of intracellular nanoparticle concentration $C''$ on the applied (intercellular) concentration $C$ can be derived as

$$C''(C) = \frac{\bar{\lambda}(C)}{V_{\text{cell}}} \approx \frac{A_1}{V_{\text{cell}}} \left(1 - e^{-k_1 C}\right) < \frac{A_1}{V_{\text{cell}}} < C \tag{18}$$

with $V_{\text{cell}}$ the mean volume per cell. The intracellular concentration $C''$ also follows the saturation behavior dominated by $\bar{\lambda}(C)$, but has a value always lower than the applied concentration $C$ regardless of co-incubation conditions.

Equivalent dose $\bar{d}_1$ is defined to account for the assumed Poisson killing per nanoparticle per cell $L_1 \sim \text{Pois}(\mu_1)$, thus to quantify the mean number of killing events per nanoparticle $\mu_1$ by the LQ model given in equation (16). Initially, the mean number of killing events per nanoparticle is intuitively assumed as a binary function $\mu_1(D, C)$ on both the applied dose $D$ and nanoparticle concentration $C$. However, it should be more explicit that the mean number of killing events $\mu_1$ per nanoparticle is dependent not directly on the nanoparticle concentration $C$, but on the concentration-dependent mean number of intracellular nanoparticles acting on a cell $\bar{\lambda}(C)$. Therefore, the binary dependency $\mu_1(D, C)$ should be more explicitly written as $\mu_1(D, \bar{\lambda}(C))$, accordingly, the equivalent dose per nanoparticle $\bar{d}_1(D, C)$ should also be more explicitly written as $\bar{d}_1(D, \bar{\lambda}(C))$.

Now consider the explicit expression of $\bar{d}_1(D, \bar{\lambda}(C))$. From the most common features of the local physical and/or biochemical effects that are all mediated via the surface of randomly agglomerated nanoparticles, it can be deduced that the increased killing events per nanoparticle per cell might be dominated by an effective surface area of a single nanoparticle, therefore the equivalent dose $\bar{d}_1(D, \bar{\lambda}(C))$, defined to account for the increased killing per nanoparticle per cell, should have an explicit dependency on an effective surface area of a single nanoparticle.

But what is the definition of the effective surface area of a single nanoparticle ? Considering that the nanoparticles acting on a cell are usually randomly agglomerated, and that the collective area of the nanoparticles' surface facing outside is thus reduced, we can first define the total effective area of all the nanoparticles per cell, $\bar{S}$, as the total average area of the exposed nanoparticle surface via which the collective effects are mediated, then we define the effective surface area per nanoparticle $\bar{s}_1$, as the equal division of the total effective area $\bar{S}$ by the mean number of nanoparticles per cell $\bar{\lambda}(C)$.

With the definition of the effective surface area per nanoparticle $\bar{s}_1$, the formulation of the single nanoparticle's equivalent dose $\bar{d}_1(D, \bar{\lambda}(C))$ can be updated, as having (1) a linear dependency on the effective area per particle $\bar{s}_1$, (2) a linear dependency on the radiation dose $D$, and (3) an independent LQ killing of all the individual nanoparticles per cell, as stated in 2.3.3. Under the assumed linear dependencies, an explicit expression of $\bar{d}_1$ can be written as

$$\bar{d}_1(D, C) = \bar{d}_1(D, \bar{\lambda}(C)) = KD\bar{s}_1\left(\bar{\lambda}(C)\right) \tag{19}$$

in which $K$ is the scaling factor, i.e. the ratio of equivalent dose $\bar{d}_1$ to the applied dose $D$ per unit effective area. By this definition, the potential nonlinear dependency of $\bar{d}_1$ on $\bar{\lambda}(C)$ is reduced to a linear dependency of $\bar{d}_1$ on $\bar{s}_1$ and a geometric dependency of $\bar{s}_1$ on $\bar{\lambda}(C)$, i.e. the effective surface area per nanoparticle on the mean number of nanoparticles per cell. It should be noted that in equation (19), the assumed linear dependency of $\bar{d}_1$ on $\bar{s}_1$ may introduce some error, due to the volume dependent local dose and radiolysis enhancements not being fully taken into account. (Potential volume dependency of $\bar{d}_1$ is discussed in section 5).

According to the definition of the effective surface area, either for all the nanoparticles or for each of the nanoparticles per cell, one can imagine that the total or single effective area should be strongly affected by the agglomeration state of the nanoparticles. Thus for different degrees of agglomeration of the nanoparticles per cell, we can make a geometric analysis on the dependencies of the total effective area $\bar{S}$, and thus the single effective area $\bar{s}_1$, on the mean number of nanoparticles per cell $\bar{\lambda}$, then by taking the derived $\bar{s}_1(\bar{\lambda})$ relation into equation (19), the binary expression of $\bar{d}_1(D, \bar{\lambda})$ can be obtained. The mean number of increased killing events per nanoparticle $\mu_1(D, \bar{\lambda})$ can thus be derived by taking $\bar{d}_1(D, \bar{\lambda})$ into equation (16), and the total killing $\bar{L}(D, \bar{\lambda})$ contributed by all the nanoparticles per cell, can be derived on the assumption of an independent additive LQ killing of $\mu_1(D, \bar{\lambda})$ for each individual nanoparticle per cell. Following this way, the expressions of the correlated quantities $\bar{\lambda} \rightarrow \bar{S} \rightarrow \bar{s}_1 \rightarrow \bar{d}_1 \rightarrow \mu_1 \rightarrow \bar{L}$ can be derived in turn for different agglomeration state of the nanoparticles per cell.

Suppose the mean geometric surface area of a single nanoparticles is $s_0$, in monodispersion state of no agglomeration, the total effective area of all the nanoparticles per cell is $\bar{S}(\bar{\lambda}) = s_0 \bar{\lambda}$ and the effective area per nanoparticle, as the equal division of $\bar{S}$ by definition, is $\bar{s}_1 = \bar{S}(\bar{\lambda})/\bar{\lambda} = s_0$. Taking this in equation (19), the equivalent dose per nanoparticle is derived as $\bar{d}_1 = KD\bar{s}_1 = KDs_0$, thus by equation (16) the increased killing events per nanoparticle per cell is $\mu_1 = \alpha(KDs_0) + \beta(KDs_0)^2$ and the increased killing by all the nanoparticles per cell can be derived as $\bar{L} = \alpha(\bar{\lambda}\bar{d}_1) + \beta(\bar{\lambda}\bar{d}_1)^2 \approx \bar{\lambda}\mu_1 = \bar{\lambda}(\alpha\bar{d}_1 + \beta\bar{d}_1^2)$ on the assumption of independent additive killing, as given in table 1(a).

In the opposite extreme case that all the nanoparticles per cell are tightly compacted into a single sphere, the total effective area is the surface area of the single sphere $\bar{S}(\bar{\lambda}) = s_0 \bar{\lambda}^{2/3}$ thus the effective surface area per nanoparticle is $\bar{s}_1(\bar{\lambda}) = \bar{S}(\bar{\lambda})/\bar{\lambda} = s_0 \bar{\lambda}^{-1/3}$. Accordingly the binary dependencies of the equivalent dose $\bar{d}_1(D, \bar{\lambda})$, the killing events per nanoparticle per cell $\mu_1(D, \bar{\lambda})$ and the total killing events $\bar{L}(D, \bar{\lambda})$ per cell, can be derived in the same way, as given in table 1(b).

From the total effective area $\bar{S}(\bar{\lambda}) = s_0\bar{\lambda}$ for monodispersed nanoparticles, and $\bar{S}(\bar{\lambda}) = s_0\bar{\lambda}^{2/3}$ for a compact sphere, a generalized expression of the total effective area can be deduced as $\bar{S}(\bar{\lambda}) = s_0\bar{\lambda}^\eta$ for the natural state of randomly agglomerated nanoparticles, with the exponent $\eta$ defined as a dispersity factor ranging from 2/3 to 1, corresponding to the increase of nanoparticle dispersity (decrease of agglomeration). If the variation of $\eta$ for different mean number of nanoparticle per cell is considered, the dispersity factor is further generalized as $2/3 < \eta(\bar{\lambda}) < 1$, thus the total effective area $\bar{S}(\bar{\lambda}) = s_0\bar{\lambda}^{\eta(\bar{\lambda})}$. The expressions of the corresponding quantities are given in table 1(c) and 1(d), respectively.

To sum up the derivation, as a theory of equivalent approximation for the enhanced cell killing in nanoparticle radiosensitization, the theoretical framework describes nanoparticles mediated dose survival modification as a consequence of compound Poisson additive killing, i.e. an independent additive radiation killing by the uniform equivalent doses of all the individual nanoparticles per cell under the LQ model, thus leading to a LQ parameter modification of the dose survival relationship. From the framework, a compound Poisson killing model is derived for an analytical prediction of the LQ parameter modification, as described in 2.4.

**Table 1.** Analytical expressions of the correlated quantities $\bar{\lambda}, \bar{S}, s_1, \bar{d}_1, \mu_1, \bar{L}$ derived for different levels of agglomeration of the nanoparticles. $\bar{\lambda}(C)$ is the concentration-dependent mean nanoparticle number per cell, $\bar{S}(\bar{\lambda})$ the total effective surface area of all the nanoparticles per cell, $\bar{s}_1(\bar{\lambda})$ the effective area per nanoparticle, $\bar{d}_1(D,\bar{\lambda})$ the binary dependency of the equivalent dose per nanoparticle, $\mu_1(D,\bar{\lambda})$ the mean number of killing events per cell by the equivalent dose $\bar{d}_1(D,\bar{\lambda})$, and $\bar{L}(D,\bar{\lambda})$ the additive killing events contributed by all the nanoparticles per cell.

a. Monodispersion: maximum dispersity ($\eta=1$)

| | |
|---|---|
| Mean nanoparticle number per cell | $\bar{\lambda}(C)=A_1(1-e^{-k_1 C})$ |
| Total effective surface area | $\bar{S}(\bar{\lambda})=s_0\bar{\lambda}$ |
| Effective area per nanoparticle | $\bar{s}_1(\bar{\lambda})=\bar{S}/\bar{\lambda}=s_0$ |
| Equivalent dose per nanoparticle | $\bar{d}_1(D,\bar{\lambda})=KD\bar{s}_1=KDs_0$ |
| Killing events per nanoparticle per cell | $\mu_1(D,\bar{\lambda})=\alpha\bar{d}_1+\beta\bar{d}_1^2=\alpha(KDs_0)+\beta(KDs_0)^2$ |
| Total additive killing per cell | $\bar{L}(D,\bar{\lambda})=\alpha(\bar{\lambda}\bar{d}_1)+\beta(\bar{\lambda}\bar{d}_1)^2\approx\bar{\lambda}\mu_1=\bar{\lambda}(\alpha\bar{d}_1+\beta\bar{d}_1^2)=\bar{\lambda}\alpha(KDs_0)+\bar{\lambda}\beta(KDs_0)^2$ |

b. Compact single sphere: minimum dispersity ($\eta=2/3$)

| | |
|---|---|
| Mean nanoparticle number per cell | $\bar{\lambda}(C)=A_1(1-e^{-k_1 C})$ |
| Total effective surface area | $\bar{S}(\bar{\lambda})=s_0\bar{\lambda}^{2/3}$ |
| Effective area per nanoparticle | $\bar{s}_1(\bar{\lambda})=\bar{S}/\bar{\lambda}=s_0\bar{\lambda}^{-1/3}$ |
| Equivalent dose per nanoparticle | $\bar{d}_1(D,\bar{\lambda})=KD\bar{s}_1=KDs_0\bar{\lambda}^{-1/3}$ |
| Killing events per nanoparticle per cell | $\mu_1(D,\bar{\lambda})=\alpha\bar{d}_1+\beta\bar{d}_1^2=\alpha(KDs_0\bar{\lambda}^{-1/3})+\beta(KDs_0\bar{\lambda}^{-1/3})^2$ |
| Total additive killing per cell | $\bar{L}(D,\bar{\lambda})=\alpha(\bar{\lambda}\bar{d}_1)+\beta(\bar{\lambda}\bar{d}_1)^2\approx\bar{\lambda}\mu_1=\bar{\lambda}(\alpha\bar{d}_1+\beta\bar{d}_1^2)=\bar{\lambda}\alpha(KDs_0\bar{\lambda}^{-1/3})+\bar{\lambda}\beta(KDs_0\bar{\lambda}^{-1/3})^2$ |

c. Random dispersion I: constant dispersity ($2/3<\eta<1$)

| | |
|---|---|
| Mean nanoparticle number per cell | $\bar{\lambda}(C)=A_1(1-e^{-k_1 C})$ |
| Total effective surface area | $\bar{S}(\bar{\lambda})=s_0\bar{\lambda}^\eta$ |
| Effective area per nanoparticle | $\bar{s}_1(\bar{\lambda})=\bar{S}/\bar{\lambda}=s_0\bar{\lambda}^{\eta-1}$ |
| Equivalent dose per nanoparticle | $\bar{d}_1(D,\bar{\lambda})=KD\bar{s}_1=KDs_0\bar{\lambda}^{\eta-1}$ |
| Killing events per nanoparticle per cell | $\mu_1(D,\bar{\lambda})=\alpha\bar{d}_1+\beta\bar{d}_1^2=\alpha(KDs_0\bar{\lambda}^{\eta-1})+\beta(KDs_0\bar{\lambda}^{\eta-1})^2$ |
| Total additive killing per cell | $\bar{L}(D,\bar{\lambda})=\alpha\bar{\lambda}\bar{d}_1+\beta(\bar{\lambda}\bar{d}_1)^2\approx\bar{\lambda}\mu_1=\bar{\lambda}(\alpha\bar{d}_1+\beta\bar{d}_1^2)=\bar{\lambda}\alpha(KDs_0\bar{\lambda}^{\eta-1})+\bar{\lambda}\beta(KDs_0\bar{\lambda}^{\eta-1})^2$ |

d. Random dispersion II: number-dependent dispersity $(2/3<\eta(\bar{\lambda})<1)$

| | |
|---|---|
| Mean nanoparticle number per cell | $\bar{\lambda}(C)=A_1(1-e^{-k_1 C})$ |
| Total effective surface area | $\bar{S}(\bar{\lambda})=s_0\bar{\lambda}^{\eta(\bar{\lambda})}$ |
| Effective area per nanoparticle | $\bar{s}_1(\bar{\lambda})=\bar{S}/\bar{\lambda}=s_0\bar{\lambda}^{\eta(\bar{\lambda})-1}$ |
| Equivalent dose per nanoparticle | $\bar{d}_1(D,\bar{\lambda})=KD\bar{s}_1=KDs_0\bar{\lambda}^{\eta(\bar{\lambda})-1}$ |
| Killing events per nanoparticle per cell | $\mu_1(D,\bar{\lambda})=\alpha\bar{d}_1+\beta\bar{d}_1^2=\alpha(KDs_0\bar{\lambda}^{\eta(\bar{\lambda})-1})+\beta(KDs_0\bar{\lambda}^{\eta(\bar{\lambda})-1})^2$ |
| Total additive killing per cell | $\bar{L}(D,\bar{\lambda})=\alpha\bar{\lambda}\bar{d}_1+\beta(\bar{\lambda}\bar{d}_1)^2\approx\bar{\lambda}\mu_1=\bar{\lambda}(\alpha\bar{d}_1+\beta\bar{d}_1^2)=\bar{\lambda}\alpha(KDs_0\bar{\lambda}^{\eta(\bar{\lambda})-1})+\bar{\lambda}\beta(KDs_0\bar{\lambda}^{\eta(\bar{\lambda})-1})^2$ |

## 2.4 Compound Poisson killing (CPK) model for nanoparticle mediated LQ parameter modification

### 2.4.1 General formulation

The mean number of nanoparticles per cell $\bar{\lambda}(C)$ can be described as a saturation function with the increase of concentration $C$, as given in equation (17), while considering the general state of number-dependent agglomeration of the nanoparticles per cell, the binary dependency of the equivalent dose per nanoparticle $\bar{d}_1(D,C)$, defined in equation (19), can be described below with the general expression of $\bar{s}_1(\bar{\lambda})$ given in table 1(d)

$$\bar{d}_1(D,C) = \bar{d}_1\left(D, \bar{\lambda}(C)\right) = KD\bar{s}_1\left(\bar{\lambda}(C)\right) = KDs_0\bar{\lambda}(C)^{\eta(\bar{\lambda}(C))-1} \qquad (20)$$

Substituting equation (20) to (16), then (16) to (15), the survival modification under the Poisson approximation of the increased killing events per cell, can be described as a consequence of independent additive LQ killing (neglecting the square cross terms) by the equivalent doses of the individual nanoparticles (of the deterministic number $\bar{\lambda}$) per cell

$$\begin{aligned} S(D,C) &= e^{-(\alpha D + \beta D^2)} e^{-\bar{\lambda}(C)\mu_1(D,C)} \\ &= e^{-(\alpha D + \beta D^2)} e^{-\bar{\lambda}(C)\left[\alpha\left(KDs_0\bar{\lambda}(C)^{\eta(\bar{\lambda}(C))-1}\right) + \beta\left(KDs_0\bar{\lambda}(C)^{\eta(\bar{\lambda}(C))-1}\right)^2\right]} \\ &= e^{-(\alpha D + \beta D^2)} e^{-(\alpha'' D + \beta'' D^2)} = e^{-\left[(\alpha+\alpha'')D + (\beta+\beta'')D^2\right]} \end{aligned} \qquad (21)$$

thus the survival modification function is

$$\gamma(D,C) = e^{-(\alpha'' D + \beta'' D^2)} = e^{-\bar{\lambda}(C)\left[\alpha\left(KDs_0\bar{\lambda}(C)^{\eta(\bar{\lambda}(C))-1}\right) + \beta\left(KDs_0\bar{\lambda}(C)^{\eta(\bar{\lambda}(C))-1}\right)^2\right]} \qquad (22)$$

It can be seen that the apparent parameters ($\alpha''$, $\beta''$) are in fact the correction values acting on the intrinsic LQ parameters ($\alpha$, $\beta$) corresponding to the survival modification by the nanoparticle mediated additive killing, and proportional to the intrinsic ($\alpha$, $\beta$) values, with the proportional coefficients (i.e. relative corrections) nonlinearly dominated by the mean number of nanoparticles per cell $\bar{\lambda}(C)$, thus nonlinearly dependent on the nanoparticle concentration $C$ as

$$\begin{cases} \dfrac{\alpha''}{\alpha} = \bar{\lambda}(C)\left(Ks_0\bar{\lambda}(C)^{\eta(\bar{\lambda}(C))-1}\right) \\ \dfrac{\beta''}{\beta} = \bar{\lambda}(C)\left(Ks_0\bar{\lambda}(C)^{\eta(\bar{\lambda}(C))-1}\right)^2 \end{cases} \qquad (23)$$

Equation (23) establishes the general formulation of the CPK model for the LQ parameter modification caused by nanoparticle radiosensitization. The physical meaning and simplification of the expressions are further discussed in 2.4.2.

### 2.4.2 Relative corrections ($\alpha''/\alpha$), ($\beta''/\beta$): physical meaning and simplification

From the definition of the equivalent dose per nanoparticle $\bar{d}_1 = KD\bar{s}_1$ (equation (19)) and the general expression of the effective surface area per nanoparticle $\bar{s}_1(\bar{\lambda}) = \bar{S}/\bar{\lambda} = s_0\bar{\lambda}^{\eta(\bar{\lambda})-1}$ (table 1(d)), we can define a dimensionless ratio $\varphi_1 = \bar{d}_1/D = Ks_0\bar{\lambda}^{\eta(\bar{\lambda})-1}$ as the induction coefficient of a single nanoparticle's equivalent dose, thus the relative corrections ($\alpha''/\alpha$),($\beta''/\beta$) in equation (23) can be rewritten as

$$\begin{cases} \dfrac{\alpha''}{\alpha} = \bar{\lambda}\dfrac{\bar{d}_1}{D} = \bar{\lambda}\varphi_1 \\ \dfrac{\beta''}{\beta} = \bar{\lambda}\dfrac{\bar{d}_1}{D}\dfrac{\bar{d}_1}{D} = \bar{\lambda}\varphi_1^2 = \left(\dfrac{\alpha''}{\alpha}\right)\varphi_1 \end{cases} \qquad (24)$$

The physical meaning of ($\alpha''/\alpha$) can thus be interpreted as the additive sum of the induction coefficient $\varphi_1$ and ($\beta''/\beta$) the additive sum of squares of the coefficient $\varphi_1^2$ over the mean number of nanoparticles per cell $\bar{\lambda}(C)$, respectively, with $\beta''/\beta = (\alpha''/\alpha)\varphi_1$. It is easy to imagine that the induction coefficient of single nanoparticle's equivalent dose has a tiny value $\varphi_1 \ll 1$ and reasonable to assume that the additive sum of the coefficients over the mean nanoparticle number per cell is still less than 1, $\alpha''/\alpha = \bar{\lambda}\varphi_1 < 1$, thus $\beta''/\beta = (\alpha''/\alpha)\varphi_1 \ll \alpha''/\alpha < 1$ is ignorable in practice. By the simplification the LQ parameter modification can be reduced to a concentration dependent correction only to the α parameter

$$\frac{\alpha''}{\alpha} = \bar{\lambda}\varphi_1 = Ks_0\bar{\lambda}(C)^{\eta(\bar{\lambda}(C))}, \quad \left(\frac{\beta''}{\beta} \approx 0\right) \tag{25}$$

It can be seen more clearly that the parameter correction ($\alpha''/\alpha$) is dominated by $\bar{\lambda}$ - the mean number of nanoparticles per cell, and affected by agglomeration through $\varphi_1$ - the agglomeration-dependent induction coefficient for the equivalent dose of single nanoparticle, with the definition $\varphi_1 = \bar{d}_1/D = Ks_0\bar{\lambda}^{\eta(\bar{\lambda})-1}$.

Equation (25) provides the first simplified expression of the CPK model for nanoparticles mediated LQ parameter modification. However, to give a practical prediction of the relative correction ($\alpha''/\alpha$), the reduced formula of equation (25) is still insufficient, because the functional relation of $\eta(\bar{\lambda})$, the parameter values of $K, s_0$ from equation (19) and $A_1, k_1$ in $\bar{\lambda}(C)$ from equation (17), are all still unknown.

The functional relation of the dispersity factor $\eta(\bar{\lambda})$ is difficult to know, due to the lack of knowledge on the number-dependent agglomeration of intracellular nanoparticles. In order to have an operable form of the CPK model, the general function $\eta(\bar{\lambda}(C))$ can be simplified as a constant value $\eta(\bar{\lambda}(C)) \approx \eta \in (2/3, 1)$, thus the concentration dependent correction ($\alpha''/\alpha$) is simplified as a linear term of the $\eta$ power of $\bar{\lambda}(C)$, and rewritten as a three-parameter dependency on concentration $C$

$$\frac{\alpha''}{\alpha} \approx Ks_0\bar{\lambda}(C)^{\eta} = Ks_0\left[A_1\left(1-e^{-k_1C}\right)\right]^{\eta} = \boldsymbol{A}(1-e^{-kC})^{\eta} \tag{26}$$
$$(\boldsymbol{A} = Ks_0 A_1^{\eta}, \quad k = k_1, \quad 2/3 < \eta < 1)$$

When the intracellular agglomeration is completely ignored, i.e. the nanoparticles per cell are assumed as ideal monodispersion with the dispersity factor $\eta=1$, the concentration dependent correction ($\alpha''/\alpha$) will be further simplified as a linear term of $\bar{\lambda}(C)$, and rewritten as a two-parameter expression on concentration $C$

$$\frac{\alpha''}{\alpha} \approx Ks_0\bar{\lambda}(C) = Ks_0 A_1\left(1-e^{-k_1C}\right) = \boldsymbol{A''}(1-e^{-kC}) \tag{27}$$
$$(\boldsymbol{A''} = Ks_0 A_1, \quad k = k_1, \quad \eta = 1)$$

Equations (26)(27) give more concise expressions of the CPK model by further simplification and parameter combination from equation (25). For practical use, a unified formula is further derived in 2.4.3, based on which two prediction models are provided.

### 2.4.3 Unified formula, prediction models and relative difference
By a transformation of equation (26), the unknown constant $\eta$ acting on $\bar{\lambda}(C)$ can be incorporated into the linear coefficient of a two-parameter expression of $(\alpha''/\alpha)^{1/\eta}$ on the concentration $C$

$$\left(\frac{\alpha''}{\alpha}\right)^{1/\eta} = (Ks_0)^{1/\eta}\bar{\lambda}(C) = (Ks_0)^{1/\eta}A_1\left(1-e^{-k_1C}\right) = \boldsymbol{A'}(1-e^{-kC}) \tag{28}$$
$$(\boldsymbol{A'} = \boldsymbol{A}^{1/\eta} = (Ks_0)^{1/\eta}A_1; \quad k = k_1; \quad 2/3 < \eta < 1)$$

Combining equation (27) into (28), a unified formula of the simplified CPK model is obtained as

$$\left(\frac{\alpha''}{\alpha}\right)^{\frac{1}{\eta}} = \boldsymbol{A'}(1-e^{-kC}), \quad (2/3 < \eta \lesssim 1) \tag{29}$$

Equation (29) gives a unified operable formula of the CPK model for making a prediction of the parameter correction of ($\alpha''/\alpha$). Note that in this unified formula, the coefficient $k$ is the rate constant in the exponential saturation function $\bar{\lambda}(C)$ inherited from equation (17), having the dimension of reciprocal concentration, while the combined coefficient $A' = (Ks_0)^{1/\eta}A_1$ from equation (28) is a non-dimensional constant not independent of, but inherently correlated with agglomeration degree through the $\eta$ value. For any specific $\eta$ value, the unknown parameter values of $(A', k)$ can be directly calculated by solving the equations of $(\alpha''/\alpha)^{1/\eta}$ on $C$ using two sets of $(\alpha''/\alpha, C)$ data measured at two concentrations $(C_1, C_2)$, then employed for an interpolation prediction of the $(\alpha''/\alpha)^{1/\eta}$ value, thus the $(\alpha''/\alpha)$ value at the intermediate concentration within $(C_1 < C < C_2)$.

In practice, through the unified equation (29), the interpolation prediction of (α''/α) can be performed using two practical models:
- *Monodispersion model*: the parameters $(A', k)$ are calculated by equation (29) using two sets of (α''/α, C) data with $\eta=1$, then directly used for interpolation prediction of the intermediate (α''/α) value at the concentration within $(C_1<C<C_2)$.
- *Agglomeration model*: the parameters $(A', k)$ are calculated by equation (29) with the same (α''/α, C) data, but giving different values for different $\eta \in (2/3,1)$. The best $\eta$ and corresponding $(A', k)$ values need to be determined by experimental data fitting at a third or more intermediate concentrations, then used for a stricter prediction of other intermediate (α''/α) value.

Basically the monodispersion model ($\eta=1$) only requires two sets of experimental (α''/α, C) data for the parameter determination, but may give an overprediction of the (α''/α) value (see below) due to complete ignorance of agglomeration. The agglomeration model may give a more accurate prediction of the (α''/α) value, but requires more experimental (α''/α, C) data points to find out the best $(A', k, \eta)$ values. A data processing scheme of the prediction models is illustrated in Appendix I.

The relative difference between the two practical models can be assessed by the concentration dependent (α''/α) value predicted for different $\eta$ values. To avoid the mathematical difficulty for a pure explicit derivation of the relative difference, two sets of (α''/α, C) data (from 3. Experimental validation) were employed for an illustration. It is seen (figure 2a) that in the concentration range between the employed data, the agglomeration model presents a lower (α''/α) curve that deviates from the monodispersion model with the increase of agglomeration (decrease of $\eta$ value). That is, the monodispersion model gives an overprediction of the nanoparticles mediated correction value (α''/α), with the deviation increasing for a higher degree of agglomeration. However, for the employed boundary data, the (α''/α) curves for $\eta \in (2/3, 1)$ also indicate an actually small range of difference. Even for the extreme agglomeration of $\eta=2/3$, the maximum difference of the monodispersion model's overprediction is still below 12% (figure 2(b)). In practice, the real difference of the predicted (α''/α) by the two models should be evaluated for the best $\eta$ value, which needs to be determined by the best fit of the predicted LQ survival curve to the measured survival data at a third or more intermediate concentrations. Further discussion on the relative difference is given in section 5.

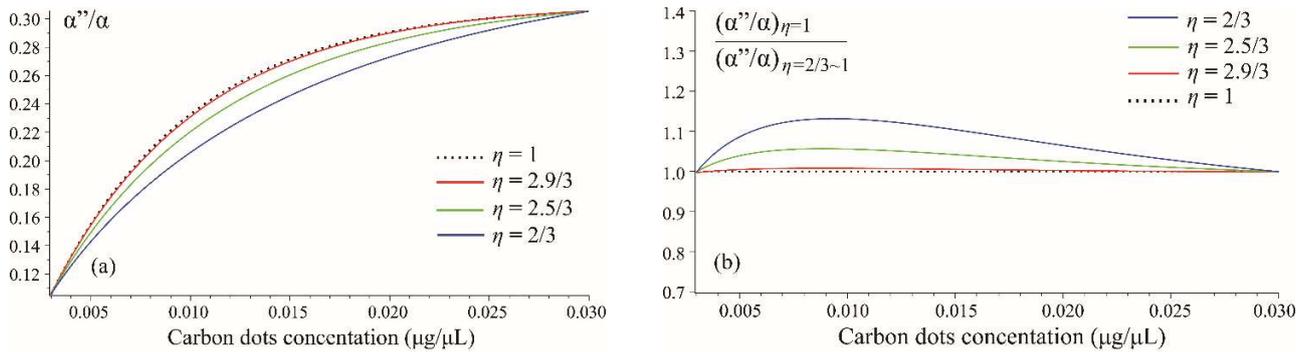

**Figure 2**. The range of variation and relative difference of the concentration dependent (α''/α) values predicted for different η. This illustration is given for a carbon dots mediated radiosensitization of HepG2 cells. Based on the boundary (α''/α, C) data ((α''/α)=0.1076 at $C$=0.003 μg/μL, (α''/α)=0.3046 at $C$=0.03 μg/μL, in Table 2), the intermediate (α''/α) value within the concentration range (0.003-0.03 μg/μL) is predicted by the CPK model for different $\eta$ values covering the possible range from 2/3 to 1, as illustrated in (a), while the relative differences of the (α''/α) value overpredicted by the monodispersion model to that predicted by the agglomeration model of different $\eta$ correction, are presented in (b).

## 3. Experimental validation

### 3.1 Cell experiments
As a tumor cell line commonly used for radiosensitivity research, human hepatocellular carcinoma (HepG2) cells were selected to make an in vitro validation of the CPK model, through the radiosensitization mediated by a common type of carbon nanodots. The Low-Z carbon dots with near-zero local dose were used for a primary validation of the model' applicability to the generally induced biochemical effects in the radiosensitization by most types of nanoparticles, in the consideration that so far few methods are available for an analytical prediction of the biochemical radiosensitization consequence. Detailed information of experiment methods on the cell culture, cytotoxicity assay, cell irradiation and clonogenic dose survival assay are provided in Appendix II.

### 3.2 Model validation

#### 3.2.1 Monodispersion model
The monodispersion model ($\eta=1$) was first validated for the prediction of the ($\alpha''/\alpha$) values in the HepG2 cells radiosensitization. From the proliferation assay, a safe range of concentration of the carbon dots (0.003-0.03 μg/μL) was selected. The interpolation prediction was performed by the boundary ($\alpha''/\alpha$) data at concentrations ($C_1$=0.003 μg/μL, $C_2$=0.03 μg/μL). Intermediate ($\alpha''/\alpha$) values at three concentrations (0.005, 0.008, 0.012 μg/μL) were predicted by equation (29) for $\eta=1$, and the corresponding LQ survival curves were compared to the clonogenic dose survival data of the sensitized HepG2 cells at the same concentrations, respectively. At each concentration the mean prediction error was calculated by

$$difference = \frac{SF_{\text{predicted}} - SF_{\text{measured}}}{SF_{\text{measured}}} \times 100 \tag{30}$$

The experimental dose survival data were obtained in the form of mean±SD, thus error propagation should also be included in the prediction error. For equations (30), the standard deviations is given by

$$\sigma(difference) = \sqrt{\frac{SF_{\text{predicted}}^2}{SF_{\text{measured}}^4} \sigma^2(SF_{\text{measured}})} \tag{31}$$

#### 3.2.2 Agglomeration correction model
The agglomeration model ($2/3<\eta<1$) was also evaluated for the prediction of the ($\alpha''/\alpha$) value in the same range of concentration (0.003-0.03 μg/μL). At the same three concentrations (0.005, 0.008, 0.012 μg/μL), the ($\alpha''/\alpha$) values were still predicted through equation (29) but using different $\eta$ value from 2/3 to 1. Some $\eta$ values greater than 1 were also tested, for potential enhancements independent of the exposed surface area of the nanoparticles per cell. (Potential volume dependency is discussed in section 5). The data fitting method for the best $\eta$ value is further described in Appendix III. As an initial validation of the correction model, the LQ dose survival data at the same concentrations were re-predicted by the best $\eta$ value, and compared to the monodispersion model on the prediction error. (Further validation on unused concentrations is still needed).

### 3.3 Results

#### 3.3.1 Radiosensitization of HepG2 cells
The cell proliferation assay showed a stable relative viability of the HepG2 cells near 0.9 (figure 3(a)) in the lower concentration range of the carbon dots below 0.016 μg/μL. With the increase of the concentration from 0.016 to 0.063 μg/μL, the relative cell viability first slightly decreased to 0.8, then sharply dropped down when the concentration exceeding 0.063 μg/μL. Accordingly, a safe concentration range of the carbon dots (0.003-0.03 μg/μL) was selected for the in vitro radiosensitization of HepG2 cells. The surviving fraction of irradiated HepG2 cells treated with different concentrations of carbon dots was determined by colony formation assay. The dose survival fraction of the carbon dots sensitized HepG2 cells at the five concentrations (0, 0.003, 0.005, 0.008, 0.012, 0.030 μg/μL) are shown in figure 3(b). It can be seen that the carbon dots led to a moderate radiosensitization of the HepG2 cells with a clear concentration dependency. At each concentration, the dose survival data were fitted to LQ model, respectively, with the finely determined LQ parameter ($\alpha$, $\beta$) values given in table 2. Moreover, the correction values of $\alpha''$ and the relative ($\alpha''/\alpha$) at the boundary concentrations (0.003, 0.03 μg/μL) were also determined for the interpolation prediction use.

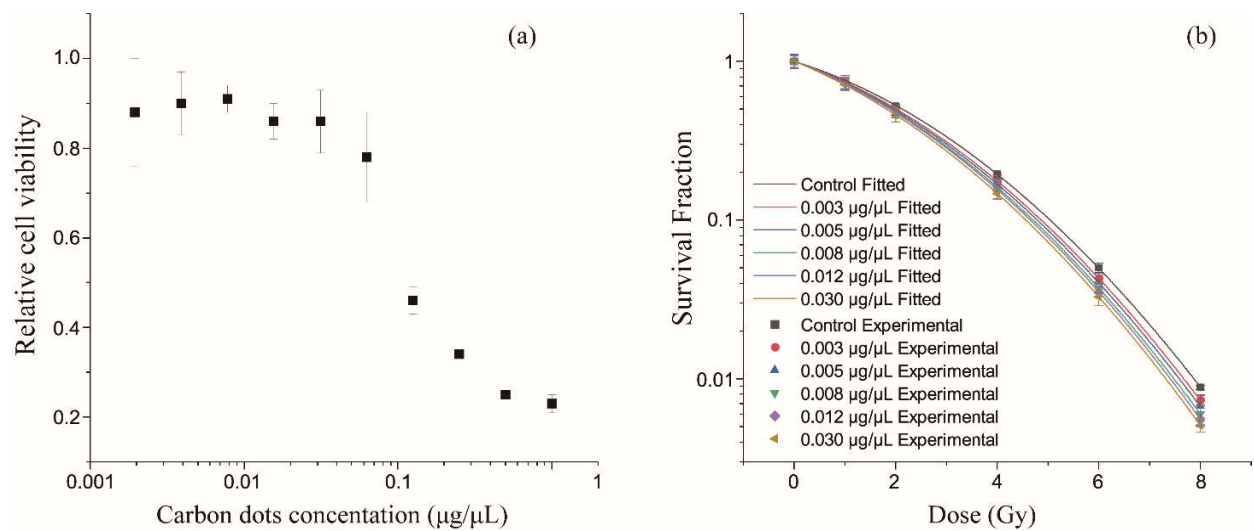

**Figure 3.** (a) Proliferation assay on the cytotoxicity and safe concentration of the carbon dots employed for the HepG2 cells radiosensitization; (b) Clonogenic dose survival of the carbon dots sensitized HepG2 cells at different concentrations, obtained under a 160 kVp X-ray irradiation.

**Table 2**. LQ parameters modification: measured for the clonogenic dose survival of HepG2 cells at different concentrations of carbon dots

| $C$ concentration (μg/μL) | α Experimental (Gy$^{-1}$) | β Experimental (Gy$^{-2}$) | α'' (mean) (Gy$^{-1}$) | (α''/α) (mean) |
|---|---|---|---|---|
| 0 (Control) | 0.22873 ± 2.02×10$^{-3}$ | 0.04523 ± 3.10×10$^{-4}$ | 0 | 0 |
| 0.003 | 0.25334 ± 1.06×10$^{-3}$ | 0.04519 ± 1.66×10$^{-4}$ | 0.02461 | 0.10759 |
| 0.005 | 0.26446 ± 1.73×10$^{-3}$ | 0.04517 ± 3.01×10$^{-4}$ | | |
| 0.008 | 0.27373 ± 3.59×10$^{-3}$ | 0.04558 ± 5.65×10$^{-4}$ | | |
| 0.012 | 0.28132 ± 3.22×10$^{-3}$ | 0.04590 ± 5.00×10$^{-4}$ | | |
| 0.030 | 0.29841 ± 1.17×10$^{-3}$ | 0.04518 ± 1.73×10$^{-4}$ | 0.06968 | 0.30464 |

**Table 3**. LQ parameter modification: predicted by the monodispersion model at the intermediate concentrations of the carbon dots

| (α''/α, C) Boundary data | Model parameters | $C$ concentration (μg/μL) | (α''/α) Predicted | α Predicted (Gy$^{-1}$) | β (Control) (Gy$^{-2}$) |
|---|---|---|---|---|---|
| (0.10759, 0.003 μg/μL) | $A'$= 0.306 | 0.005 | 0.15757 | 0.26477 | 0.04523 |
| (0.30464, 0.030 μg/μL) | $k$=142.727 (μg/μL)$^{-1}$ | 0.008 | 0.21029 | 0.27683 | 0.04523 |
| | $\eta$=1 | 0.012 | 0.25318 | 0.28664 | 0.04523 |

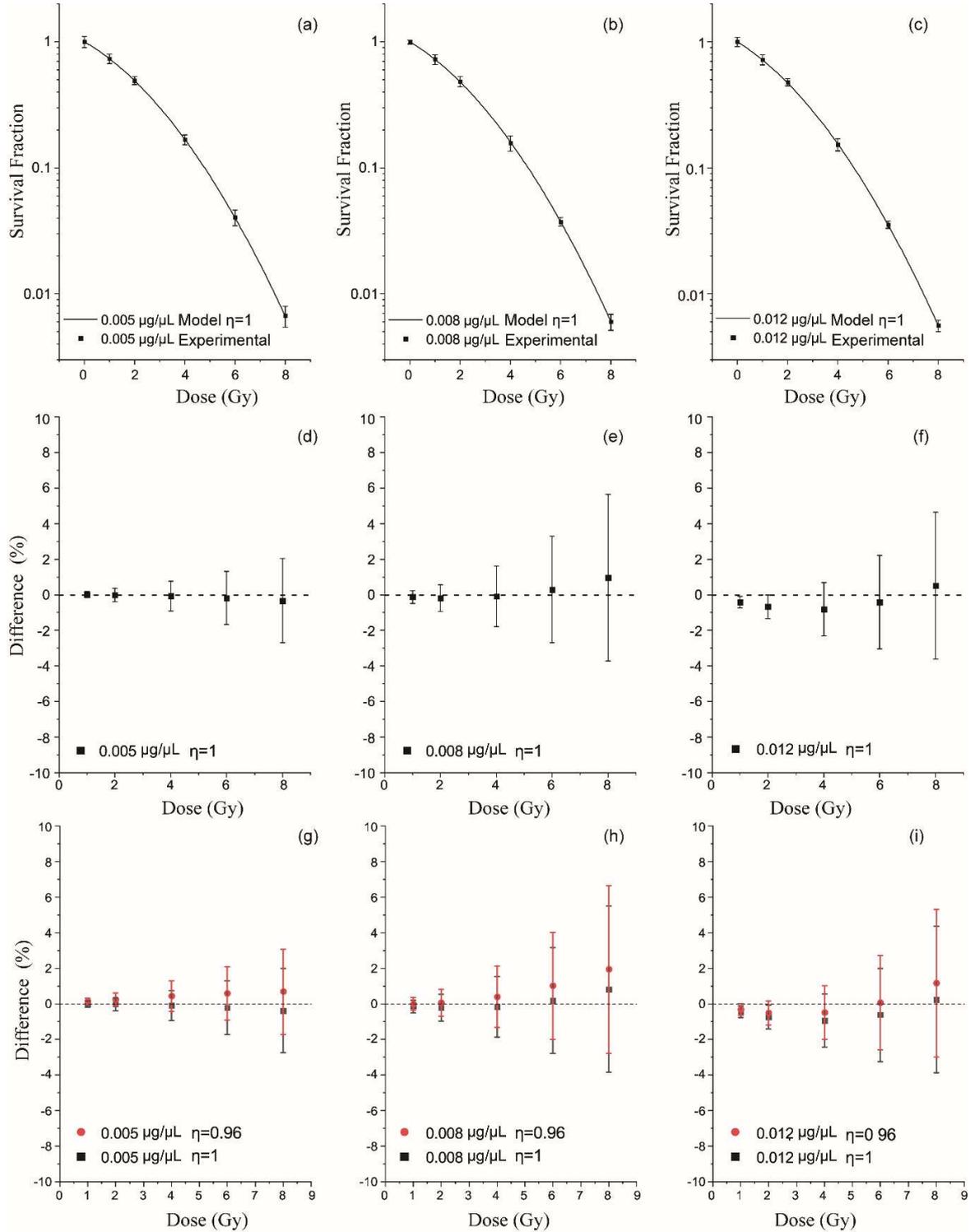

**Figure 4**. An in vitro validation of the CPK model by the radosensitization of HepG2 cells by a common type of carbon dots. (a-c) Comparison of the LQ survival curve (solid line) predicted by the monodispersion model ($\eta=1$) with the clonogenic dose survivals (solid square), measured at three intermediate concentrations (0.005, 0.008, 0.012 μg/μL) of the carbon dots within the range (0.003-0.03 μg/μL), by a 160 kVp X-ray irradiation at a dose rate of 0.75 Gy/min. (d-f) The prediction error of the monodispersion model relative to the clonogenic dose survival data, as the percentage difference calculated by equations (30)(31), indicates a low prediction error less than 2%. (g-i) Comparison of the prediction errors of the agglomeration model ($\eta=0.96$) and the monodispersion model ($\eta=1$), relative to the experimental data of clonogenic dose survival.

*3.3.2 Validation of the prediction models*

With the experimental (α"/α) data determined at the two boundary concentrations (table 2), the (α"/α) values at the intermediate concentrations (0.005, 0.008, 0.012 μg/μL) were first predicted by the monodispersion model (equation (29) for η=1), with the parameters (A', k) and the predicted LQ parameters (α, β) given in table 3. The LQ curves corresponding to the predicted (α, β) values are plotted in figure 4(a)(b)(c), in comparison with the clonogenic dose survival data measured at the three concentrations. At each concentration, the prediction error was calculated by equations (30),(31), and presented in figure 4(d)(e)(f), respectively. It can be seen that the monodispersion model gave an accurate prediction of the intermediate (α"/α) values, with the mean error of the corresponding LQ curves less than 2% relative to the clonogenic dose survival data.

For different $\eta$ values, the agglomeration correction model (equation (29) for 2/3<η<1) gave different (α"/α), thus different LQ survival curves at the same concentrations. The goodness of fit of the predicted LQ survivals to the measured data was evaluated by the adjusted coefficient of determination $R^2_{adj}$ by equations (A.1),(A.2) (More details in Appendix III), with the $\eta$-dependent variation of the $R^2_{adj}$ value shown in figure A.2, from which the best $\eta$ value was determined at around $\eta \approx 0.96$. Corresponding to the (α"/α) values re-predicted for $\eta$=0.96, the LQ survival curves at the same three intermediate concentrations were re-compared with the clonogenic dose survival data, with the prediction error shown in figure 4(g)(h)(i) together with that of the monodispersion model. However, further reduction of the prediction error not observed, even higher fluctuation appeared. It is attributed to the insignificant variation of the $R^2_{adj}$ value ($\approx$1) (figure A.2) that indicates a negligible agglomeration effect. In other words, for the radiosensitization of HepG2 cell by the carbon dots, the relative difference of the monodispersion model and agglomeration correction model is so small, that the LQ parameter modification can be predicted with sufficiency accuracy by the monodispersion model, which describes the carbon dots mediated radiosensitization as purely additive killing effect that linearly dominated by the mean number of nanoparticles per cell.

## 4. Discussion

The CPK model is derived to make an analytical description, instead of local effect modeling, for nanoparticle radiosensitization. The theoretical framework is based on a series of assumptions, simplifications or approximations, summarized as:
- Starting from classic target theory, thus taking no account of the intercellular bystander effects and/or non-targeting effects;
- Assuming an independent occurrence of intrinsic radiation killing and nanoparticle mediated killing events (equation (7),(8));
- Two Poisson quantities assumed: number of nanoparticles per cell $Z$, killing event number per nanoparticle $L_1$ (equation (9));
- Independent occurrence of the killing events per nanoparticle, leading to a compound Poisson number of the additive killing events by the Poisson number of nanoparticles per cell (equations (10)-(12));
- Using the deterministic mean number $Z=\bar{\lambda}(C)$ instead of the Poisson number $Z \sim \text{Pois}(\bar{\lambda}(C))$ of the nanoparticles per cell to simplify the compound Poisson additive killing (equation (10)) into a Poisson killing (equations (13)-(15));
- The concept of single nanoparticle's equivalent dose $\bar{d}_1$ (equation (16)) has three implicit assumptions: (1) The biochemical killing contributed by each nanoparticle is equated to a Poisson dose killing following the LQ model, (2) the subcellular local physical/biochemical killing is equated to a global uniform dose killing and (3) the collective killing by multiple nanoparticles is equated to an independent additive killing by multiple exposures to the single nanoparticle's equivalent dose.
- The concentration dependency of $\bar{\lambda}(C)$ is approximated as a negative exponential saturation function (equation (17));
- The nonlinear dependency of $\bar{d}_1$ on $\bar{\lambda}$ is simplified as a linear dependency of $\bar{d}_1$ on $\bar{s}_1$(single nanoparticle's equivalent surface area) and a nonlinear dependency of $\bar{s}_1$ on $\bar{\lambda}$ (equation (19)). Substituting the expressions of $\bar{\lambda}(C)$ and $\bar{d}_1(D,\bar{\lambda}(C))$ (table 1) into equation (16),(15) gives the general formulation of nanoparticles mediated dose survival modification equations (20-22) (Potential dependency of $\bar{d}_1$ on nanoparticle volume, other than the surface area dependency on $\bar{s}_1$, is further discussed below).

The above steps lead to a general form of the CPK model (equation (23)) for nanoparticles mediated LQ parameter modification. For practical use, simplified forms of the model are further derived:
- Neglecting the β correction reduces the CPK model to a LQ parameter modification only to the α parameter (equation (25));
- Assuming a constant dispersity factor $\eta(\bar{\lambda}(C)) \approx \eta \in (2/3,1)$ gives a simplified expression of (α"/α) as a linear term of $\eta$ power of the nanoparticle number per cell $\bar{\lambda}(C)$ and organized as a three parameter dependency on the concentration (equation (26));
- Assuming $\eta$=1 gives a linear function of (α"/α) on $\bar{\lambda}(C)$, and written as a two-parameter (α"/α)-C dependency (equation (27));
- From equations (26)(27), a unified equation is derived as an operable formula of the model for prediction use (equation (29)).

Note that the monodispersion and agglomeration models based on equation (29) are both proposed for interpolation prediction of the concentration dependent parameter correction of (α"/α), and only applicable to a safe range of nanoparticle concentration. Comparative data (not shown) indicated that the model is not suitable for extrapolation prediction of the (α"/α) value at a higher or lower concentration outside the boundary, due to the increase of prediction error with the deviation out of the interval range.

From the carbon dots validation (figure 4), it can be seen that the monodispersion model gave such a sufficiently high accuracy in the interpolation prediction of the (α"/α) value that little improvement can be made by the agglomeration model's η correction. Therefore, for practical use of the CPK model, two questions need to be further discussed:

- *Is the η correction necessary ?*

Theoretically it can be judged by the range of relative difference of the (α"/α)-C curves predicted for different η value ($2/3<\eta\leq1$). As illustrated in figure 2a, the η correction leads to a reduction of the (α"/α) value relative to monodispersion model's prediction, but mostly within 10%, thus for the employed boundary data, in most cases, an interpolation prediction by the monodispersion model is accurate enough, i.e. the carbon dots radiosensitization can be regarded as a purely additive killing effect, with the LQ parameter correction (α"/α) linearly dominated by the mean number of nanoparticles per cell $\bar{\lambda}(C)$.

Mathematically, for the interpolation prediction by equation (29), the range of variation of the intermediate (α"/α)-C curves for different $\eta\in(2/3<\eta\leq1)$ are purely determined by the boundary ((α"/α), C) data, thus for a given range of concentration, one can directly determine the range of variation of the intermediate (α"/α) value by the boundary data and ignore the $\eta$ correction when the maximum difference of the (α"/α) values (for η=2/3 and η=1) is small enough. Otherwise when the maximum difference is so large that a monodispersion model cannot ensure the accuracy, the $\eta$ correction is still needed.

The mathematical judgement raises a new question: is the $\eta$-correction negligible in most case in nanoparticle radiosensitization? If this is the case, the monodispersion model is sufficient for a common use. But due to an explicit form of the relative difference still being unavailable, for different materials and cells, the applicability of monodispersion model still needs specific validation by the mathematical judgement via specific boundary ((α"/α), C) data.

- *How to make the η correction ?*

The $\eta$ correction is to find the $\eta$ value which best reflects the concentration dependency of the (α"/α) value under equation (29). For that, more experimental data at intermediate concentrations are needed for a function fitting to equation (29) in combination with the boundary data. Testing results (not shown) indicated that, a single intermediate ((α"/α), C) data point is insufficient to give a consistent fitting of the $\eta$ value. In principle, the more data points employed, the better consistent $\eta$ value can be obtained, but at the expense of the model's predictive capability due to overreliance on experimental data. As a compromise, at least two or three intermediate ((α"/α), C) data points are basically required, to ensure a sufficient consistency in the fitting determination of the best $\eta$ value.

The function fitting is performed by adjusting the $\eta$ value to acquire the highest goodness of fit of equation (29) to the employed multiple ((α"/α), C) data points. The goodness of fit is evaluated by the adjusted coefficient of determination $R^2_{adj}$ described in Appendix III. Note that the experiment data employed for the $\eta$ fitting are actually multiple dose-survival data points at different concentrations, thus in equation (A.1) for $R^2_{adj}$ calculation, the sample size $n$ is the total number of dose survival data points at different concentrations, while the number of explanatory variables (independent variables that affecting the cell survival) $p=2$ when the survival data are treated as the binary function value of $S(D,C)$. When the $\eta$-dependent variation of the $R^2_{adj}$ value is determined, its range of variation will indicate whether the $\eta$ correction is really necessary.

Strict validation of the agglomeration model should be made at the intermediate concentrations other than that already employed in the fitting of the best $\eta$ value. However, in this work, the initial validation of the $\eta$ correction was still performed at the same three intermediate concentrations already used in the $\eta$ fitting (figure 4 (g)(h)(i)), thus in theory, reduced prediction error should be a corollary, but the difference is actually too small to be discriminated from the prediction error of the monodispersion model (This is also reflected by the insignificant $\eta$-dependent variation of the $R^2_{adj}$ value in figure A.2), therefore further validation of the agglomeration model is still necessary in future work.

It should be pointed out that as an analytical description model for nanoparticle radiosensitization, the CPK model is constructed through a general derivation on the basic assumption of a compound Poisson additive killing, therefore theoretically, the model is expected to be generally applicable to the radiosensitization effects by most types of nanoparticles. Due to the topic focusing on analytical derivation of the the model and the limitation on article length, the current validation is only presented for a single cell line sensitized by a single type of nanoparticles. The low-Z carbon nanodots of no local dose effect are not commonly used in nanoparticle radiosensitization, but may serve as a single effect sensitizer, for a primary validation on the general biochemical effects in the radiosensitization by most types of nanoparticles, while the local dose (and radiolysis) effects can be added in the model (with potential modification, see below) for high-Z nanoparticles based on the biochemical effects. More comprehensive validation of the model for different types of nanoparticles and cell lines will be reported in future studies.

Finally, some further discussion is necessary on the definition of single nanoparticle's equivalent dose $\bar{d}_1$. In current framework, the equivalent dose per nanoparticle is assumed to be linearly related to the nanoparticle's effective surface area (equation (19)). However, the assumption for surface effects is only true for the effects catalyzed by biochemical reactions on the surface of the nanoparticles. For the local dose and radiolysis yield enhancements, often contributed by high-Z nanoparticles, the sensitization is dependent upon the probability of interactions between the incident radiation and the nanoparticle. The interaction probability depends on the amount of material available to interact within and is hence volume dependent. Therefore, the linear dependency of $\bar{d}_1$ on $\bar{s}_1$ may introduce some error due to the volume effects not being fully included. The fact that the surface model in this work fits the carbon dots radiosensitization quite well, is likely due to the enhancement effects primarily via surface chemistry, with limited contribution of local dose and radiolysis yield enhancements.

Meanwhile, it is also interesting to note that, for the value of $\eta$ that reflects the effect of agglomeration, there may be a potential case of "over-dispersion", with the best correction value $\eta>1$. This is a purely theoretical speculation but may be the truth when nanoparticle radiosensitization is not merely mediated by surface effects, but is affected, even dominated by certain non-surface effects that lead to an apparent $\eta$ value greater than 1. Clearly, for high-Z nanoparticles, such a potential requirement for higher $\eta$ values may arise, and likely be due to the volume effects of local dose and radiolysis yield. Therefore, for high-Z nanoparticles of greater local effects, it is vital in future work to test whether a volume dependency term of $\bar{d}_1$ should be included in the model.

In future study, further extension and validation will be carried out to fully understand the applicability of CPK model, including:
- A stricter carbon dots validation of the agglomeration model, for the same HepG2 cells and the same 160 kVp X-ray radiation, but for the intermediate data other than that already employed in the fitting determination of the best $\eta$ value.
- Extended carbon dots validation of the model (monodispersion model for $\eta=1$ and agglomeration model for $2/3<\eta<1$ through equation (29)) for more different cell lines (cancerous, normal) and radiations (photons, protons, heavy ions).
- Extending the definition of single nanoparticle's equivalent dose $\bar{d}_1$ to account for both surface and volume dependent killing enhancements, thus improving the accuracy of the model for more different types of nanoparticles.
- Extended validation of the model for four types of nanoparticles (low-Z non-metallic, high-Z metallic, with or without surface modification enhanced killing). Emphasis will be put on the scope of application (1) from the carbon dots to other unmodified low-Z nanoparticles by which the radiosensitization is often dominated by an enhanced oxidative stress, (2) from low-Z non-metallic to unmodified high-Z nanoparticles of which the radiosensitization effects are dominated by local dose and radiolysis enhancements, and (3) from unmodified nanoparticles to functionalized nanoparticles of specific modifications (for targeting, drug loading, chemo synergy or sensitization, imaging enhancement or other multi-functionalization) that lead to or help for an enhanced radiation cell killing.
- Concluding the applicability of the CPK model for nanoparticle radiosensitization for different materials, cells and radiations.
- Concluding the applicability of the monodispersion model in the applicable scope of the CPK model, i.e. whether it is enough accurate to use the monodispersion model in most practical cases of nanoparticle radiosensitization.

## 5. Conclusion

To construct an analytical framework instead of local effect modeling for an easier prediction of nanoparticle radiosensitization, a survival modification framework is derived on the assumption of a nanoparticles mediated compound Poisson additive killing, which describes the nanoparticles mediated dose survival modification as a consequence of independent additive killing by the individual nanoparticles of Poisson number per cell, through a uniform equivalent dose per nanoparticle under LQ killing model. Based on the framework a compound Poisson killing model is constructed for an analytical prediction of nanoparticles mediated LQ parameter modification, with two prediction models (monodispersion model and agglomeration correction model) proposed for different states of agglomeration ($\eta=1$ and $2/3<\eta<1$) of the nanoparticles per cell. In vitro validation by the radiosensitization of HepG2 cells by carbon dots indicated a high accuracy of the model, with the relative errors of predicted dose survival curves both within 2% by the two prediction models. Though further improvements are still needed, the presented model of compound Poisson additive killing provides a promising method for an analytical prediction of the biological effectiveness of nanoparticle radiosensitization, and easier to use for the potential application in the treatment planning of nanoparticle enhanced radiotherapy. Extension of the theoretical framework and further validation on the applicability to different cells, radiations and nanoparticles, will be carried out to in future work.

**Acknowledgment**
The work was supported by National Natural Science Foundation of China (No. 11375047, 11005019, 12075063) and Shanghai Scientific Research Project (18441905600).

# Appendix

## I: Data processing scheme for interpolation prediction

A data processing scheme of the CPK model is illustrated in figure A.1. Interpolation prediction of the concentration dependent ($\alpha''/\alpha$) value is made through the unified formula (equation (29)) using the boundary ($\alpha''/\alpha, C$) data. The monodispersion model uses $\eta=1$ for a direct prediction of the intermediate ($\alpha''/\alpha$) value by the boundary data, while the agglomeration model makes a fitting determination of the best $\eta$ value at first, by changing the $\eta$ value to fit equation (29) to a selected number of experimental intermediate ($\alpha''/\alpha$) data (in combination with the boundary data) to find out the $\eta$ value that giving the highest goodness of fit, then uses the best $\eta$ value for the interpolation prediction of the intermediate ($\alpha''/\alpha$) values, with a potentially improved accuracy.

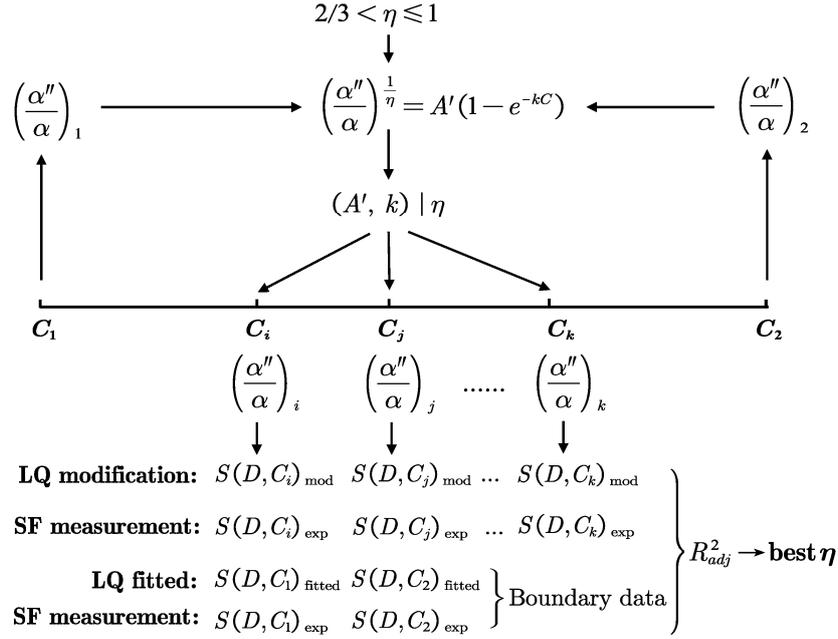

**Figure A.1.** Data processing scheme of the CPK model for interpolation prediction of nanoparticles mediated LQ parameter modification.

## II: Cell experiment methods

### A.1.1 Cell culture

The HepG2 cells (Shanghai Cell Bank of Chinese Academy of Sciences, China) were cultured in DMEM medium (Gibco, MD, USA) containing 10% of fetal bovine serum (Gemini Foundation, USA) and 1% of Penicillin-Streptomycin solution (Beyotime Biotechnology, China) in a humidified incubator, at 37°C with 5% CO2.

### A.1.2 Carbon dots cytotoxicity assay

The carbon dots (Jienasi New Materials, Nanjing, China) employed for the CPK model validation are a common type of amino modified carbon nanodots (3-5 nm), with positive surface charge in the aqueous phase, and an initial concentration of 10 mg/mL. According to the manufacturer recommended range of concentration for in vitro use, cytotoxicity assay of the carbon dots was performed for the serial dilutions from 1 μg/μL down to 9 lower concentrations (0.500, 0.250, 0.125, 0.063, 0.032, 0.016, 0.008, 0.004, 0.002 μg/μL). HepG2 cells were seeded in 96-well plates in a density of 15000 per well and divided into 11 groups, for a co-incubation with the carbon dots of the different concentrations (including 0 μg/μL as the control group). A blank group of pure medium (no cells, no carbon dots) was also set to eliminate the background effect. After the first 12-hr incubation in well, the culture medium of the cells was replaced by the carbon dots of the 11 concentrations, respectively, for 3 days co-incubation. Subsequently, 10 μL of CCK-8 solution (Dojindo, Kumamoto, Japan) was added in each well and the plate put back in incubator for 2 hrs. After that the optical density (OD) in each well was measured in a microplate reader (MultiSkan FC, Thermo Scientific) at the wavelength 450 nm. The relative cell viability can be determined by [(OD$_{experimental}$ − OD$_{blank}$)/(OD$_{control}$ − OD$_{blank}$)]×100%.

*A.1.3 Cell irradiation and clonogenic dose survival assay*

Single cell suspension of the HepG2 cells was prepared by a centrifugation of exponential phase cells at 2000 rpm for 5 minutes, then plated into 60 mm dishes at different seeding densities appropriate for different irradiation dose. Each group of cells for a specific radiation dose was pre-treated by the carbon dots at different concentrations (0, 0.003, 0.005, 0.008, 0.012, 0.03 μg/μL). The dishes of zero concentration were set as control group. After 24 hrs co-incubation with carbon dots, nonadherent cells were discarded, and the adherent cells were carefully washed twice in phosphate buffered saline (PBS) and added with fresh medium. After that the cell irradiation was performed under a 160 kVp X-ray irradiator (IXS 1650, VJ Technologies (Suzhou) Inc, China) by different doses (0, 1, 2, 4, 6, 8 Gy) at a dose rate 0.75 Gy/min that calibrated by a X-ray dosimeter (QUART didoEASY R, QUART GmbH, Germany) in the center of the irradiation platform. The irradiated cells were further incubated to form colonies, with the medium replaced every 3-4 days for about 14 days. The formed colonies were fixed by methanol and stained with 0.1% crystal violet solution (Beijing Solarbio Science technology, China). Finally, the number of colony was counted in a microscope, with one colony having at least 50 cells. All the experiments were repeated for at least three times. The clonogenic dose survival data were normalized to the control data based on the plating efficiency and given as mean±SD. All the data were fitted to LQ model with the (α, β) parameter values carefully determined. In addition, the ratio of cell survival at the boundary concentrations (0.003, 0.03 μg/μL) to the survival data of control group was fitted to $\gamma(D,C) \approx e^{-\alpha'' D}$ (by equation (21),(22),(25)), respectively, to determine the correction values of α", thus the relative values of (α"/α) for the interpolation prediction use in model validation.

**III: Data fitting for the best η value**

For the validation of the agglomeration model, the *η* value that best reflects the agglomeration level needs to be first determined In this work, different *η* values were used to fit the LQ curves predicted by equation (29) to the clonogenic survival data at the intermediate concentrations of the carbon dots. The goodness of fit of the predicted LQ survival curve to the clonogenic survival data was calculated by the adjusted coefficient of determination

$$R_{adj}^2 = 1 - (1-R^2)\frac{(n-1)}{n-p-1} \tag{A.1}$$

in which *n* is the sample size (number of measured/predicted data points) in the calculation of $R^2$ (in below), while *p* is the total number of explanatory variables (independent variables that affecting the cell survival) in the model. In this work, *n*=30 for the dose survival data under six doses for the three concentrations together with the boundary data at the two concentrations, while *p*=2 for the data under the binary dependency $S(D,C)$. The coefficient of determination $R^2$ is defined as

$$R^2 \equiv 1 - \frac{SS_{\text{res}}}{SS_{\text{tot}}}; \quad \left(SS_{\text{res}} = \sum_{i=1}^{n}(y_i - f_i)^2, \ SS_{\text{tot}} = \sum_{i=1}^{n}(y_i - \overline{y})^2, \ \overline{y} = \frac{1}{n}\sum_{i=1}^{n} y_i\right) \tag{A.2}$$

Here $SS_{\text{res}}$ is the residual sum of squares, $SS_{\text{tot}}$ is the total sum of squares, $y_i$ is the measured and $f_i$ is the predicted survival data at different doses and concentrations of the total number *n*, $\overline{y}$ is the mean of the measured data.

The *η* value that leads to the highest $R_{adj}^2$ was defined as the best *η* value that reflects the agglomeration correction to the (α"/α). For the radiosensitization of HepG2 cell by the carbon dots in the present work (section 3.3.1), the *η*-dependent variation of the $R_{adj}^2$ value calculated by equations (A.1)(A.2) is shown in figure A.2, by which the best *η* value was determined around *η*≈0.96.

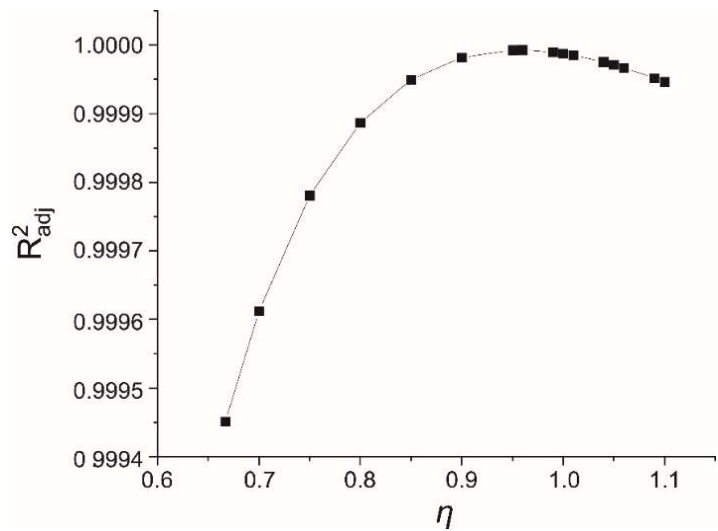

**Figure A. 2**. Goodness of fit of the agglomeration model to the experimental dose survival for different $\eta$ value. The agglomeration dependent variation of the $R^2_{adj}$, calculated for the LQ survival curves predicted under different $\eta$ value, indicated the best $\eta$ value for the highest $R^2_{adj}$.